\documentstyle[12pt]{article} \begin{document}\thispagestyle{empty}
\begin{center} \LARGE \tt \bf {Resonant amplification of magnetic field seeds and early universe dynamos from chiral torsion currents}
\end{center}

\vspace{1.0cm}

\begin{center}
{\large By L.C. Garcia de Andrade\footnote{Departamento de F\'{\i}sica Te\'{o}rica - IF - UERJ - Rua S\~{a}o Francisco Xavier 524, Rio de Janeiro, RJ, Maracan\~{a}, CEP:20550.e-mail:garcia@dft.if.uerj.br}}
\end{center}

\begin{abstract} Earlier Kostelecky [Phys Rev D 69, 105009 (2004)] has investigated the role of gravitational sector in Riemann-Cartan spacetime (RC) with torsion, in Lorentz and CPT violating (LV) standard model extension (SME). In his paper use of QED extension in RC spacetime is made. More recently the author [Phys Lett B (2011)] obtained magnetic field galactic dynamo seeds in the bosonic sector with massless photons, which proved to decay faster than the necessary [Phys Lett B (2012)] to be able to seed galactic dynasmos. In this report it is shown that by using the fermionic sector of Kostelecky Lagrangean and torsion written as a chiral current, one obtains torsion and magnetic fields explicitly from a Heisenberg-Ivanenko form of Dirac equation whose solution allows us to express torsion in terms of LV coefficients and magnetic field in terms of fermionic matter fields. When minimal coupling between electromagnetic and torsion fields is used it is shown that the fermionic sector of QED with torsion leads to resonantly amplify magnetic fields which mimics an ${\alpha}^{2}$-dynamo mechanism. Fine tunning of torsion is shown to result in the dynamo reversal, a phenomenon so important in solar physics and geophysics. Of course this is only an analogy since torsion is very weak in solar and geophysics contexts. An analogous expression for the ${\alpha}-effect$ of mean field dynamos is also obtained where the alpha effect is mimic by torsion. Similar resonant amplification mechanisms connected to early universe have been considering by Tsagas and Finelli et al.
\end{abstract}
\newpage

\section{Introduction}
 Primordial magnetic fields (PMF) and LV have been recently investigated by Bertolami et al \cite{1} in the context of Riemannian geometries of string theories, where galactic dynamo seeds could be obtained. Some other attemps of obtaning such PMF using new physics in the sense of QED and generalised Maxwell theories in the backgrounds of other alternative gravity theories have been followed \cite{2}. More recently we \cite{2} have tried to show that masssless photon-torsion coupling in the realm of quantum electrodynamics (QED) could reach this task. In that paper flat torsion modes, where by flat we consider Minkowski spacetime ${\textbf{M}}^{4}$ plus torsion are obtained. Unfortunatly more recent we show \cite{3} that the magnetic field decayed fast enough not to be able to seed galactic dynamos. De Sabbata and Sivaram \cite{4} have investigated in detail the Heisenberg-Ivanenko spinor electrodynamics. In section 2 we investigate how this equation can be used to generate cosmic magnetic field. More recently Barrow et al \cite{5} have obtained in general relativistic cosmologies using conventional Maxwell equations, more stringent limits of $10^{-20}G$ and $10^{-12}G$. In this paper we present two examples where torsion is considered as a fermionic chiral current and magnetic field is expressed either in terms of torsion or in terms of LV coefficients. In the first case the coupling between electrodynamics and torsion is minimal, while in the second is non-minimal. This paper is organised as follows: Section 2 assumes a non-minimal photon-torsion coupling which yields a ${\alpha}^{2}$-dynamos, where an analogous mean field dynamo relation between electric and magnetic fields is obtained similar to recent results by Semikos and Sokoloff \cite{6} using electroweak phase transitions.  These fermionic currents with torsion which are able to produce cosmic magnetic fields amplification by resonance. In section 3 one makes use of the Kostelecky fermionic sector Lagrangean to obtain axial torsion in terms of the LV coefficients whose expression is quite and remarkable similar to the expression using the ${\alpha}^{2}$-dynamos. Discussions and conclusions are presented in section 4, where comparison with ${\alpha}^{2}$-dynamos with axionic torsion \cite{7}.

\section{Dynamo mechanism from Heisenberg-Ivanenko-Cartan electrodynamics}
Earlier Forbes et al \cite{8} have shown that QCD domain walls with spinor fields and spin polarised nucleons orthogonal to the wall may give rise to primordial magnetic fields. It happens that as is well-known \cite{4} the spinor currents are connected to torsion fields naturally. With this motivation in this section we consider the electrodynamical field equations with torsion and spinor currents, to show that magnetic fields maybe amplified in spacetimes with torsion and massless photons by resonance. To simplify matters we consider that the torsion modes here, are flat endowed with torsion, in the sense that the Minkowski background metric is used instead of the Friedmann spacetime used in the last section. Let us start by considering the Maxwell-Dirac-Cartan electrodynamic equations in the form
\begin{equation}
{\partial}_{\nu}F^{\rho\nu}+ \frac{1}{2}F_{\mu\nu}T^{{\mu}{\nu}{\rho}}=J^{\rho}
\label{1}
\end{equation}
where $J^{\rho}$ is the fermionic current given in terms of Dirac spinor ${\psi}$ by
\begin{equation}
J^{\rho}=g\bar{\psi}{\gamma}^{\rho}{\psi}
\label{2}
\end{equation}
By Fourier transforming equation (\ref{1}) we obtain
\begin{equation}
({\omega}^{2}-k^{2})A^{\rho}+{A}_{[{\mu},\nu]}T^{[\mu\nu\rho]}=J^{\rho}
\label{3}
\end{equation}
where
\begin{equation}
T^{[\mu\nu\rho]}={\epsilon}^{\mu\nu\rho\sigma}T_{\sigma}={\epsilon}^{\mu\nu\rho\sigma}[\bar{\psi}{\gamma}^{5}
{\gamma}_{\sigma}{\psi}]
\label{4}
\end{equation}
is the totally skew-symmetric Cartan torsion, and $T^{\mu}$ is the axial torsion. This is a chiral fermionic current. Let us now find a solution of this equation where $\textbf{k}.\textbf{A}=0$ which may be fulfilled by the relations $\textbf{A}=(A^{1},A^{2},0)$ and $\textbf{k}=(0,0,k_{3})$. From these coordinates one may decompose the Maxwell-Dirac-Cartan equation as
\begin{equation}
({\omega}^{2}-k^{2})A^{0}+A_{[0,1]}L=J^{0}
\label{5}
\end{equation}
where $J^{0}=\bar{\psi}{\gamma}^{0}{\psi}$ and $L=T^{[010]}$ is the $(0,1,0)$ component of torsion. Here also $A^{0}={\phi}$ is the electric potential. The remaining equations are collectively written as
\begin{equation}
({\omega}^{2}-k^{2})\textbf{A}+A_{[0,1]}\textbf{T}=\textbf{J}=g\bar{\psi}\vec{\gamma}\psi
\label{6}
\end{equation}
where $\textbf{T}:=T^{[01i]}$ where $(i=1,2,3)$ and the Dirac 3-D matrices are given by $\vec{\gamma}$. From these expressions one obtains after a straightforward algebra
\begin{equation}
({\omega}^{2}-k^{2})A^{2}+E_{1}M=J^{2}
\label{7}
\end{equation}
\begin{equation}
E_{1}=\frac{J^{3}}{N}
\label{8}
\end{equation}
\begin{equation}
{\phi}=\frac{g\bar{\psi}{\gamma}^{0}\psi}{({\omega}^{2}-k^{2})}
\label{9}
\end{equation}
and
\begin{equation}
A^{1}=\frac{g(\bar{\psi}{\gamma}^{1}\psi)}{({\omega}^{2}-k^{2})}
\label{10}
\end{equation}
Expression (\ref{8}) shows that the component $E_{1}=A_{[0,1]}$ of the electric field is damped by the component N of torsion. From this same expression one notices that the electric field is greatly amplified when resonance approaches at ${\omega}^{2}\approx{k^{2}}$. From expression (\ref{10}) one observes that that this same resonance amplification happens to the magnetic field since their components are
\begin{equation}
B_{1}= -ik^{3}A^{2}
\label{11}
\end{equation}
\begin{equation}
B_{2}= ik^{3}A^{1}
\label{12}
\end{equation}
where $M:=T^{[012]}$ and $N=T^{[013]}$. Substitution of (\ref{8}) into expression (\ref{7}) one obtains
\begin{equation}
({\omega}^{2}-k^{2})A^{2}=[J^{2}-J^{3}\frac{M}{N}]
\label{13}
\end{equation}
therefore a fine-tunning on torsion given by M=N off resonance yields a cut-off in the component $A^{2}$ of magnetic vector potential, and from expression (\ref{9}) yields a cut-off on the component of magnetic field itself when $J^{2}=J^{3}$. A simple expression for the magnetic field can be obtained as
\begin{equation}
B_{1}=\frac{ik_{1}(J^{2}-\frac{M}{N}J^{3})}{({\omega}^{2}-k^{2})}
\label{14}
\end{equation}
Thus this expression shows that off the fine tunning of torsion $M=N$ the magnetic field is strongly amplified by the resonance. Therefore torsion seems to be a mechanism by which one can shut off the dynamo or even enhances the onset of dynamos by the condition $M<<N$ and the constraint on torsion $M>>N$ may even induces a reversal in the magnetic field so common in solar and terrestrial dynamos. Another important role for the fine tunning of torsion here is that when we write expression (\ref{14}) in terms of the electric field $E_{1}$

\begin{equation}
B_{1}=\frac{ik_{1}(J^{2}-{M}E_{1})}{({\omega}^{2}-k^{2})}
\label{15}
\end{equation}
In this expression one notices that when one $ME_{1}>>J^{2}$ this expression reduces to $E_{1}={\alpha}B_{1}$ where ${\alpha}^{-1}:=\frac{-ik_{1}M}{{\omega}^{2}-k^{2}}$ which is exactly the expression of mean field dynamo and the $\alpha-effect$ so important in galactic dynamos. Note that ${\alpha}$ is non-singular at the resonance. Let us now write the Heisenberg-Ivanenko nonlinear spinor equation
\begin{equation}
i{\gamma}^{\mu}{\partial}_{\mu}\psi+{L_{0}}^{2}(\bar{\psi}{\gamma}^{5}{\gamma}^{\mu}{\psi})(\bar{\psi}{\gamma}^{5}
{\gamma}_{\mu}\psi)=0
\label{16}
\end{equation}
In terms of chiral torsion current this can be obtained as
\begin{equation}
i{\gamma}^{\mu}{\partial}_{\mu}\psi+{L_{0}}^{2}T^{2}=0
\label{17}
\end{equation}
By partial derivation of expression (\ref{10}) one obtains
\begin{equation}
{\partial}_{2}A_{1}=g\frac{(\bar{\psi}{\gamma}_{1}{\partial}_{2}\psi+{\partial}_{2}\bar{\psi}{\gamma}_{1}{\psi})}
{{\omega}^{2}-k^{2}}
\label{18}
\end{equation}
substitution of the expression
\begin{equation}
{\gamma}^{2}{\partial}_{2}\psi=4i{L_{0}}^{2}T^{2}
\label{19}
\end{equation}
into (\ref{18}) yields
\begin{equation}
B^{3}=-ig\frac{{L_{0}}^{2}T^{2}[\bar{\psi}{\sigma}_{12}\psi]}{{\omega}^{2}-k^{2}}
\label{20}
\end{equation}
Here ${\sigma}_{12}=({\gamma}_{1}{\gamma}_{2}-{\gamma}_{2}{\gamma}_{1})$ is the Clifford algebra. This expression from the primordial magnetic field is similar to the one obtained by Forbes et al \cite{8} in the case of PMF obtained from QCD domain walls where the spin of the nucleons are orthogonal to the wall, this physically is reasonable since in the Einstein-Cartan theory of gravity the high density of the spin of the nucleons should be proportional to torsion of spacetime, so our last expression seems to corroborate this fact through the term $T^{2}$. One notes from the last expression that the strenght of the magnetic field depends upon the fermion coupling g.

 \section{LV and PMF from chiral torsion currents}
   In this section we shall consider the solution of the spinor field equations obtained from a particular case of general Kostelecky Lagrangean density restricted to the fermionic sector of QED given by
\begin{equation}
{\cal{L}}_{K}=ek_{\alpha\beta\gamma}T^{\alpha\beta\gamma}\bar{\psi}{\psi}-(c_{eff})_{\mu\nu}\bar{\psi}{\gamma}^{\mu}
{\partial}^{\nu}{\psi}-(b_{eff})_{\mu}\bar{\psi}{\gamma}^{5}{\gamma}^{\mu}\psi+J^{\mu}A_{\mu}-
\frac{1}{4}F^{\mu\nu}F_{\mu\nu}
\label{21}
\end{equation}
where $c_{eff}$ and $b_{eff}$ are coefficients given by
\begin{equation}
({c_{eff}})_{\mu\nu}=c_{\mu\nu}-\frac{1}{2}h_{\mu\nu}+{\chi}_{\mu\nu}
\label{22}
\end{equation}
and
\begin{equation}
{(b_{eff})}_{\mu}=b_{\mu}-\frac{1}{4}{\epsilon}_{\alpha\beta\gamma\mu}{\partial}^{\alpha}{\chi}^{\beta\gamma}+\frac{1}{8}T^{\alpha\beta\gamma}{\epsilon}_{\alpha\beta\gamma\mu}
\label{23}
\end{equation}
where here k coefficient is the LV coefficient and since we are in the Minkowski spacetime the metric gravitational perturbations $h_{\mu\nu}$ and ${\chi}_{\mu\nu}$ vanish. Here $c_{\mu\nu}$ plays the role of the Minkowski metric of flat spacetime. Thus the above coefficients reduce to

\begin{equation}
({c_{eff}})_{\mu\nu}=c_{\mu\nu}
\label{24}
\end{equation}
and
\begin{equation}
{(b_{eff})}_{\mu}=b_{\mu}+\frac{1}{8}T^{\alpha\beta\gamma}{\epsilon}_{\alpha\beta\gamma\mu}
\label{25}
\end{equation}
Variation of this lagrangean in terms of the vector magnetic potential $A_{\mu}$ yields
\begin{equation}
{\partial}_{\mu}F^{\mu\nu}=J^{\nu}\sim{g\bar{\psi}{\gamma}^{\nu}\psi}
\label{26}
\end{equation}
The new Minkowskian coefficients with torsion, the Lagrangean expression above is reduced to
\begin{equation}
{\cal{L}}_{K}=[ek_{\alpha\beta\gamma}{\epsilon}^{\alpha\beta\gamma\mu}\bar{\psi}\psi-\frac{1}{8}T^{\mu}]T_{\mu}-i
\bar{\psi}{\gamma}^{\mu}{\partial}_{\mu}\psi-\frac{1}{4}F^{\mu\nu}F_{\mu\nu}
\label{27}
\end{equation}
Now let us perform the variation of the axial torsion ${\delta}T^{\mu}$ to obtain the last field equation for torsion
\begin{equation}
T^{\mu}=8ek_{\alpha\beta\gamma}{\epsilon}^{\alpha\beta\gamma\mu}\bar{\psi}\psi
\label{28}
\end{equation}
where in these last computations we have used the expression for torsion of the chiral fermionic current of last section. This last expression shows that the torsion is proportional to the LV coefficient k as shown by Kostelecky.  However last expression has not been explicitly found by Kostelecky. The Heisenberg-Ivanenko equation here is obtaining by variation of Kostelecky Lagrangean by performing the variation of spinor field ${\delta}\bar{\psi}$ which yields
\begin{equation}
-i{\gamma}^{\mu}{\partial}_{\mu}\psi+e(k_{\alpha\beta\gamma}{\epsilon}^{\alpha\beta\gamma\mu}T_{\mu}-[b_{\mu}+
\frac{1}{8}T_{\mu}]
{\gamma}^{5}{\gamma}^{\mu}){\psi}=0
\label{29}
\end{equation}
After some algebra one obtains the more general expression
\begin{equation}
{F^{\nu}}_{\beta}{\gamma}^{\beta}=\frac{-ieb_{\mu}}{({\omega}^{2}-k^{2})}[\bar{\psi}{\gamma}^{5}{\sigma}^{\mu\nu}{\psi}]
\label{30}
\end{equation}
Decomposition of this equation yields the electric field
\begin{equation}
\vec{\gamma}.\textbf{E}=\frac{-ie\textbf{b}}{({\omega}^{2}-k^{2})}[\bar{\psi}{\gamma}^{5}{\vec{\sigma}}^{0}{\psi}]
\label{31}
\end{equation}
\begin{equation}
\textbf{E}{\gamma}^{0}+\vec{\gamma}\times{\textbf{B}}=\frac{-ieb_{i}}{({\omega}^{2}-k^{2})}[\bar{\psi}{\gamma}^{5}{\vec{\sigma}}^{i}{\psi}]
\label{32}
\end{equation}
\section{Primordial magnetic fields from Dirac fields}
In this section we shall be concerned with the simpler case of considering the non-back reaction of the Ivanenko-Heisenberg equation which are simpler to solve. Solution for the magnetic field are simpler than the ones obtained by Forbes et al \cite{8} however the magnetic field $B^{3}$ of section 2, is not amplified at resonance by the simple fact that we just consider Dirac fields that are homogeneous in the isotropic universe. Let us consider the fully Heisenberg-Ivanenko equation (\ref{29}) above in the form
\begin{equation}
-i{\gamma}^{0}{\partial}_{t}\psi-e(kT_{0}-[b_{0}+\frac{1}{8}T_{0}]{\gamma}^{5}{\gamma}^{0}){\psi}=0
\label{33}
\end{equation}
Let us consider the free-particle solution approximation
\begin{equation}
\psi\sim{{\psi}_{(0)}exp{i{\omega}t}}
\label{34}
\end{equation}
Substitution of this expression into equation (\ref{33}) yields the constraint equation
\begin{equation}
-i{\omega}{\gamma}^{0}-e[kT_{0}\textbf{1}-(b_{0}+\frac{1}{8}T_{0}){\gamma}^{5}{\gamma}^{0}]=0
\label{35}
\end{equation}
where $\textbf{1}$ is the unit matrix. Expanding the Dirac matrices one obtains
\begin{equation}
{\omega}=ekT_{0}=-8ekb_{0}
\label{36}
\end{equation}
which yields the following four-dimensional spinors
\begin{equation}
\psi={\psi}_{(0)}exp[iekT_{0}t]
\label{37}
\end{equation}
Substitution of this type of Dirac spinor wave functions into the expression for the primordial magnetic field of section 1 above one obtains
\begin{equation}
B^{3}\sim{gL_{0}T^{2}\frac{({\bar{\psi}}_{0}{\sigma}_{3}\textbf{1}{\psi}_{0})}{\omega}}
\label{38}
\end{equation}
where we have used the identity ${\sigma}_{1}{\sigma}_{2}=i{\sigma}_{3}$ among Pauli matrices ${\sigma}_{j}$. Note that unfortunatly this field is extremely weak since the torsion still appears squared. However if one considers the expression (\ref{31}) in a highly conducting universe where the electric field vanishes and this expression implies that $\textbf{b}$ vanishes and expression (\ref{32}) reduces to
\begin{equation}
\vec{\gamma}\times{\textbf{B}}=\frac{eb_{0}}{({\omega}^{2}-k^{2})}[{\bar{\psi}}_{(0)}{\sigma}_{3}\textbf{1}{\psi}_{(0)}]
\label{39}
\end{equation}
Since now the LV factor $b_{0}$ is linear in torsion $T_{0}$ this expression for the primordial magnetic field does not have a so small contribution of torsion as before. Thus we reach the conclusion that the backreaction nonlinear terms in torsion does not contribute significantly to the primordial magnetic fields, while when chiral torsion currents are neglected torsion contributes more to the magnetic field. A simpler lagrangean where we do not consider the LV factor $b_{\mu}$ can be given by
\begin{equation}
{\cal{L}}\sim{i\bar{\psi}[{{\partial}}+{L_{0}}^{2}{\gamma}^{5}{\gamma}^{\mu}T_{\mu}-m_{N}e^{i{\phi}{\gamma}^{5}}]\psi}
\label{40}
\end{equation}
This Lagrangean when added to Maxwell Lagrangean not only reproduces the above equations of section 2 with chiral currents but also is able to reproduce the QCD domain wall of Forbes et al to introduce torsion actually this can be done simply by substituting ${\partial}_{i}-{\sigma}_{i}\rightarrow{{\partial}_{i}-{\sigma}_{i}-S_{i}}$ in terms of Pauli matrices. In the next section we shall solve this Dirac equation in the presence of torsion.
\section{Dirac and primordial magnetic fields in QCD domain walls}
In this section we shall solve the equations of QCD domain walls in the presence of torsion, in two dimensions given by
\begin{equation}
2i[{\partial}_{0}+{\sigma}_{j}{\partial}_{j}+i{L_{0}}^{2}{\sigma}_{j}T_{j}]{\Psi}_{-}=2m_{N}e^{i{\phi{\gamma}^{5}}}{\Psi}_{+}
\label{41}
\end{equation}

\begin{equation}
2i[{\partial}_{0}-{\sigma}_{j}{\partial}_{j}-i{L_{0}}^{2}{\sigma}_{j}T_{j}]{\Psi}_{+}=2m_{N}e^{-i{\phi{\gamma}^{5}}}{\Psi}_{-}
\label{42}
\end{equation}
where $m_{N}$ is the mass of the nucleons domain wall where spins are polarised and orthogonal to the wall. The Dirac $2-spinors$ components of positive helicity are given by
\begin{equation}
{{\Psi}_{+}}_{1}=\frac{1}{\sqrt{S}}{\epsilon}_{1}
\label{43}
\end{equation}
\begin{equation}
{{\Psi}_{+}}_{2}=\frac{1}{\sqrt{S}}{\epsilon}_{2}
\label{44}
\end{equation}
\begin{equation}
{{\Psi}_{-}}_{1}=\frac{1}{\sqrt{S}}{\chi}_{1}
\label{45}
\end{equation}
\begin{equation}
{{\Psi}_{-}}_{2}=\frac{1}{\sqrt{S}}{\chi}_{2}
\label{46}
\end{equation}
where S is the area of the domain wall and ${\phi}(z)$ is the domain wall solution \cite{8}. In the domain wall ${\partial}_{x}={\partial}_{y}=0$, thus the spinorial equations reduce to
\begin{equation}
-m_{N}e^{i\phi}{\chi}_{1}+i[{\partial}_{t}+{\partial}_{z}+i{L_{0}}^{2}T_{z}]{\epsilon}_{1}=0
\label{47}
\end{equation}
\begin{equation}
i[{\partial}_{t}-{\partial}_{z}-i{L_{0}}^{2}T_{z}]{\chi}_{1}-m_{N}e^{-i\phi}{\epsilon}_{1}=0
\label{48}
\end{equation}
and
\begin{equation}
-m_{N}e^{i\phi}{\chi}_{2}+i[{\partial}_{t}+{\partial}_{z}+i{L_{0}}^{2}T_{z}]{\epsilon}_{2}=0
\label{49}
\end{equation}
\begin{equation}
i[{\partial}_{t}-{\partial}_{z}-i{L_{0}}^{2}T_{z}]{\chi}_{2}-m_{N}e^{-i\phi}{\epsilon}_{2}=0
\label{50}
\end{equation}
Recalling that these equations are complex one obtains the following solution
\begin{equation}
{{\Psi}_{+}}_{1}={{\Psi}_{+}}_{2}=\frac{1}{2\sqrt{S}}e^{-{L_{0}}^{2}T_{3}cot{\phi}(t+z)}
\label{51}
\end{equation}
\begin{equation}
{{\Psi}_{-}}_{1}={{\Psi}_{-}}_{2}=\frac{1}{2\sqrt{S}}e^{{L_{0}}^{2}T_{3}cot{\phi}(t+z)}
\label{52}
\end{equation}
one notes here that the fermion helicity spinor depends upon the sign of torsion as we had stated before. Note that torsion contributes to negative helicity Dirac modes in order to amplify these modes. The important issue now is to know if these amplification of Dirac spinors implies amplification of primordial magnetic field seeds which are able to seed galactic dynamos. This is actually the case, since the Maxwell tensor $F_{{\mu}{\nu}}$ as seen from above can be expressed as seen from above can be expressed in terms of 2-spinors. The four-Dirac spinor can be expressed as
\begin{equation}
{{\Psi}}_{1}={{\Psi}}_{2}=\frac{1}{2\sqrt{S}}cosh[m_{N}{L_{0}}^{2}T_{3}ctg{\phi}(t+z)]
\label{53}
\end{equation}
\begin{equation}
{\Psi}_{3}={{\Psi}_{4}}=\frac{1}{2\sqrt{S}}sinh[{m_{N}{L_{0}}^{2}T_{3}cot{\phi}(t+z)}]
\label{54}
\end{equation}
Thus the four spinors present a dissipation due to torsion.
\section{Discussions and Conclusions}
Earlier de Forbes et al \cite{8} have investigated two dimensional Dirac spinor equations to investigate QCD domain walls and its use to investigate primordial magnetic fields which may be seeds of galactic dynamos. Their QCD domain walls align electric and magnetic dipoles orthogonal to the wall, the spin of the nucleons being similarly polarised as the ones that generates torsion in the realm of Einstein-Cartan gravity \cite{4}. This paper shows that a more general solution using a spacetime background torsion lagrangean given by Kostelecky \cite{9} in spinorial form one may obtain more general solutions than the Forbes et al \cite{8} without constraining the solutions to a QCD domain wall. When torsion is not written in the form of a chiral current model simplifies to a simple Dirac equation which we solve for homogeneous Dirac fields. Homogeneous Dirac fields allows us to obtain non-resonant primordial magnetic fields where when back-reaction non-linear fields are neglected, the galactic dynamo seeds from torsion linear modes. It is also shown that the fine-tunning of the torsion parameter $K=M-N$ above, either amplification or damping contributions of the magnetic field may be possible. Resonance amplification of the magnetic fields can be enhanced or damped by this torsion parameter. Other ways of obtaining the amplification of magnetic fields without dynamos have discussed recently \cite{10}. It seems that axial torsion considered above also appears naturally in QCD domain walls investigated by Forbes et al \cite{8}. This deserves more study and consideration in near future. After I finished the paper I noticed that a very similar relation to ours above for the ${\alpha}^{2}$ dynamo was obtained by Semikoz and Sokoloff \cite{6} were obtained from neutrino currents and weak interactions as ${\textbf{E}}_{weak}=-{\alpha}\textbf{B}$ where ${\textbf{E}}_{weak}$ is the weak interactions contribution for the electric field. The resemblance with their work and the last section of this paper becomes more clearly when one recalls that the neutrino currents that generates the ${\alpha}$ effect are spinorial fermionic currents as the ones considered in section 3. Also recently the author \cite{7} has been able to show that torsion LV bounds from CP violation and ${\alpha}^{2}-dynamos$ can be obtained in analogy to what happens in the previous reference if one considers a plasma of axionic-photon type. Finelli and Gruppuso \cite{11} have discussed previously a resonant amplification of magnetic fields and gauge fields in general, however they did not addressed the torsion problem or the domain wall problem.
\section{Acknowledgements}
    We would like to express my gratitude to D Sokoloff and A Brandenburg for helpful discussions on the subject of this paper. Special thanks go to my colleague Vitor Lemes for enlightening discussions on Dirac fields and LV. Thanks are also due to Alan Kostelecky for his constant and kind advice on LV. Financial support from CNPq. and University of State of Rio de Janeiro (UERJ) are grateful acknowledged.

\end{document}